\documentclass[journal,final,10pt,oneside,twocolumn]{IEEEtranTCOM}

\usepackage{url}
\usepackage{graphicx}
\usepackage{amsmath,bbm}
\usepackage{amssymb}
\usepackage{esint}
\usepackage{color}
\usepackage{babel}
\usepackage{graphics}
\usepackage{algorithm}
\usepackage{algpseudocode}
\usepackage{float}
\usepackage{cite}
\usepackage{mleftright}
\mleftright

\usepackage{mathtools}
\DeclarePairedDelimiterX{\infdivx}[2]{[}{]}{
  #1\,\delimsize\|\,#2%
}
\newcommand{\KLdiv}{D\infdivx}

\newcounter{algsubstate}
\renewcommand{\thealgsubstate}{\alph{algsubstate}}
\newenvironment{algsubstates}
  {\setcounter{algsubstate}{0}
   \renewcommand{\State}{
     \stepcounter{algsubstate}
     \Statex {\footnotesize\thealgsubstate:}\space}}
  {}

\newcommand{\Tr}{\textrm{Tr}}
\newcommand{\ket}[2][]{{|#2\rangle_{#1}}}
\newcommand{\bra}[2][]{{}_{#1}\langle #2|}
\newcommand{\eqnref}[1]{(\ref{#1})}
\newcommand{\figref}[1]{Fig.~\ref{#1}}


\ifCLASSINFOpdf

\else

\fi

\begin{document}

\title{Capacity of a Lossy Photon Channel With Direct Detection}

\author{Karol~\L{}ukanowski and 
        Marcin~Jarzyna
\thanks{The authors are with the Centre for Quantum Optical Technologies, Centre of New Technologies, University of Warsaw, Banacha 2c, 02-097 Warszawa, Poland. K.\L{}. is also with the Faculty of Physics, University of Warsaw, Pasteura 5, 02-093 Warszawa, Poland}
}

\markboth{IEEE Transactions on Communications}
{Submitted paper}

\maketitle

\begin{abstract}

We calculate numerically the capacity of a lossy photon channel assuming photon number resolving detection at the output. We consider scenarios of input Fock and coherent states ensembles and show that the latter always exhibits worse performance than the former. We obtain capacity of a discrete-time Poisson channel as a limiting behavior of the Fock states ensemble capacity. We show also that in the regime of a moderate number of photons and low losses the Fock states ensemble with direct detection is beneficial with respect to capacity limits achievable with quadrature detection.
\end{abstract}


\section{Introduction}

Every communication protocol is carried out by performing physical measurements on physical objects which are fundamentally described by the laws of quantum mechanics. For optical communication this means that in a regime in which signal strength is low, on the level of few photons per channel use, in order to analyze the communication performance one has to properly utilize the quantum description of light, which, at the fundamental level, is composed of photons. In particular, it is necessary to consider quantum states of light and quantum measurements \cite{Caves1994, Banaszek2020}. A standard optical channel may be modeled by a Gaussian bosonic quantum channel \cite{Shapiro2009}. The ultimate information rates for such channels, known as classical channel capacities, have been studied extensively in the literature \cite{Giovannetti2004, Giovannetti2014, Holevo2001}. However, even though the optimal ensembles of quantum states saturating the classical capacity bound for Gaussian channels have been discovered, the necessary measurement schemes remain largely unknown, with the exception of a few particular scenarios \cite{Guha2011, Jarzyna2019, Ding2019}. These involve regimes of very weak and very strong signal strengths, quantified by the average number of photons per channel use received at the output, in which respectively a photon number resolving (PNR) or a heterodyne detection become almost optimal \cite{Giovannetti2004, Guha2011, Jarzyna2019}.

On the other hand it is known that under an average power constraint for an ideal optical channel the optimal ensemble of states is given by either a Gaussian coherent states ensemble or a thermal Fock states ensemble \cite{Yuen1993, Giovannetti2004}. Importantly, the latter is composed of states $\ket{k}$ of exactly $k$ photons which can transmit information encoded only in the energy. One may show that for such an ensemble the optimal measurement, saturating the classical capacity bound, is given by a PNR measurement. This is not the case for the coherent states ensemble as they are reminiscent of classical states of light and encode information not only in their energy but also in the phase of the optical field. Since a PNR measurement allows one to detect the number of photons in the signal, an optical channel utilizing PNR detection at the output is known as a photon channel.

Despite the same performance for a perfect scenario, when one introduces losses into the channel, the Fock states ensemble no longer remains optimal, whereas the coherent states ensemble still allows to attain the classical capacity bound with appropriately modified output mean photon number. However, as in the ideal case, a physical realization of the optimal measurement for the latter is unknown. For the Fock states ensemble, on the other hand, the PNR measurement is always optimal since they can carry information encoded only in the Fock state number. One may show that this strategy allows to obtain at most half of the classical capacity offered by the coherent states ensemble in the large power limit \cite{Bowen1968}. At first glance it seems therefore that the ability to transmit information encoded in phase is crucial for optimal performance, even if one deals with very small losses. However, since the Fock states ensemble strategy approaches the classical capacity in the weak signal regime \cite{Bowen1968} it may perform well also for moderate powers. In particular, it is interesting how it compares to protocols based on quadrature or PNR detection performed on coherent states ensembles, the latter resulting in what is known as a discrete-time Poisson channel \cite{Shamai1990}.

In this paper we investigate the capacity of a lossy optical channel without any kind of additive noise with PNR detection at the output which together forms a so-called lossy photon channel. We specifically consider input ensembles composed of Fock and coherent states. In order to find the capacity we employ the Blahut-Arimoto algorithm \cite{Blahut1972, Arimoto1972} which allows us to obtain also the optimal prior probability distributions for both types of states. We recover results obtained earlier in \cite{Gordon1962, Bowen1968, Martinez2007} in the asymptotic limit of large signal strength quantified by the average number of photons per channel use $\bar{n}$, while also optimizing the information rate numerically for finite $\bar{n}$. We show that the capacity attainable with the Fock states ensemble in the limit of low channel transmission while keeping the received average number of photons constant is equal to the capacity of a discrete-time Poisson channel with the optimal prior distributions being the same for both types of channels.

\section{Information theory}

A classical communication channel is described by a set of conditional probability distributions $p(y|x)$ which characterize how output symbols $y$ depend on the input $x$. In a quantum picture \cite{NielsenChuang, Wilde2017, Desurvire2009} this probability distribution arises through the Born rule $p(y|x)=\Tr \left[\Lambda\left(\rho_x\right)\Pi_y\right]$, where $\rho_x$ denotes a, possibly mixed, quantum state encoding an input symbol $x$, $\Lambda$ is a completely positive trace preserving map known as quantum channel describing the evolution of states through the channel and $\Pi_y$ is the positive operator-valued measure (POVM) characterizing the measurement performed by the receiver. Assuming respective input states are sent with prior probability $p(x)$ the maximal information rate per channel use $R$ is given by the mutual information
\begin{equation}
R\leq I(X:Y)=H(Y)-H(Y|X),
\label{eq:mutual}
\end{equation}
where $H(Y)=-{\sum_y p(y)\log_2 p(y)}$ and $H(Y|X)=-{\sum_{x,y}p(x)p(y|x)\log_2 p(y|x)}$ are Shannon entropy and conditional entropy. Mutual information optimized over the prior probability distribution is known as capacity
\begin{equation}\label{eq:capacity}
C=\max_{\{p(x)\}} I(X:Y),
\end{equation}
which quantifies the best communication performance utilizing a certain set of states $\left\{\rho_x\right\}$ and a POVM $\left\{\Pi_y\right\}$. In order to further maximize the communication rate one can optimize \eqnref{eq:capacity} over the ensemble of input states $\left\{\rho_x\right\}$ or the POVM $\left\{\Pi_y\right\}$. Crucially, the latter optimization needs to include in principle collective measurements over an arbitrarily large number of channel outputs, which makes the problem formidable. Fortunately, it can be shown \cite{Holevo1973, Holevo1998, Schumacher1997} that a saturable upper bound on the transmission rate is given by the classical channel capacity
\begin{equation}
C_{\textrm{class}}=\max_{\{\rho_x, p(x)\}}\left\{S\left[\Lambda\left(\bar{\rho}\right)\right]-\sum_x p(x) S\left[\Lambda\left(\rho_x\right)\right]\right\},
\label{eq:class_capacity}
\end{equation}
where $S(\rho)=-{\Tr\left(\rho\log_2\rho\right)}$ denotes the von Neumann entropy and $\bar{\rho}=\sum_x p(x) \rho_x$ is the average state. Note that in \eqnref{eq:class_capacity} one still needs to perform optimization over the input ensemble, whereas the optimization with respect to POVMs is already included. However, the tools used in the proof of \eqnref{eq:class_capacity} do not allow to find any practical detection schemes saturating the bound \eqnref{eq:class_capacity} in a straightforward way. The resulting POVM may be highly unfeasible, with implementation requiring a simultaneous collective measurement on a large number of time slots. This means that even if one finds a solution to \eqnref{eq:class_capacity}, i.e., the optimal ensemble of input states, in order to find a corresponding measurement that would be also practical one still needs to refer to \eqnref{eq:mutual} and perform optimization over POVMs by hand, which is usually mathematically or numerically intractable.

Importantly, for optical channels one needs to specify some constraints on the input signal, otherwise the information rate may become infinite. A common choice is to fix an average signal optical power $P$. For a signal with a central carrier frequency $f_c$ and a bandwidth $B$ this constraint is equivalent to fixing the average number $\bar{n}$ of signal photons per time bin, as the latter is equal to $\bar{n}=P/(2\pi B\hslash f_c)$, where $h$ is the Planck's constant. Other constraints are also viable, e.g., a constraint on the maximal signal power in the link, but we will not consider them here.

\section{Lossy bosonic quantum channel}

A basic example of a quantum optical communication channel for which the information rate is known is the lossy channel \cite{Giovannetti2004, Holevo2001}. For such a channel, assuming a given signal average number of photons $\bar{n}$, the classical capacity \eqnref{eq:class_capacity} reads
\begin{equation}\label{eq:ideal}
C_{\textrm{class}}=g(\eta\bar{n}),\quad g(x)=(x+1)\log_2(x+1)-x\log_2 x,
\end{equation}
where the function $g(x)$ denotes the von Neumann entropy of a thermal state with an average number of photons $x$ and $\eta$ denotes the optical transmission of the channel. 

A particular case of $\eta=1$ corresponds to an ideal noiseless optical channel. As mentioned earlier, for such a case there exist two optimal ensembles of input states known to saturate \eqnref{eq:ideal}: a Gaussian coherent states ensemble
\begin{equation}\label{eq:coherent_ens}
\rho_\alpha=\ket{\alpha}\bra{\alpha},\quad p(\alpha)=\frac{1}{\pi \bar{n}}e^{-|\alpha|^2/\bar{n}},
\end{equation}
where the amplitude $\alpha\in\mathbb{C}$, and a thermal Fock states ensemble
\begin{equation}\label{eq:Fock_id}
\rho_n=\ket{n}\bra{n},\quad p(n)=\frac{\bar{n}^n}{(\bar{n}+1)^{n+1}}.
\end{equation}
Importantly, coherent states are often regarded as classical as they represent quantum states of light emitted by a well stabilized laser and they can be represented as a superposition of Fock states $\ket{\alpha}=e^{-|\alpha|^2/2}\sum_{n=0}^\infty\frac{\alpha^n}{\sqrt{n!}}\ket{n}$ \cite{Glauber1963, Sudarshan1963}.
The average state $\bar{\rho}$ of both ensembles \eqnref{eq:coherent_ens} and \eqnref{eq:Fock_id} is a thermal state with an average number of photons $\bar{n}$. Since in the absence of losses the output states of both ensembles are pure, the only contribution to the capacity in \eqnref{eq:class_capacity} is given by the entropy of the average state and equal to \eqnref{eq:ideal}. A practical scheme saturating the bound in \eqnref{eq:ideal}, however, is known only for the latter ensemble and is given by the PNR measurement.

Introducing losses into the channel, i.e., taking $\eta<1$, corresponds to decreasing the information rate. Interestingly, in this case, even for infinitesimally small losses, only the coherent states ensemble remains optimal and saturates the bound in \eqnref{eq:ideal}. This is because coherent states remain pure under the influence of a lossy channel, only their amplitude decreases $\ket{\alpha}\to\ket{\sqrt{\eta}\alpha}$. Therefore, the only contribution to the information rate is given by the average state which, at the channel output, is equal to a thermal state with an average number of photons $\eta\bar{n}$. On the other hand, Fock states become mixed after propagation through a lossy channel
\begin{equation}\label{eq:fock_lossy}
\ket{k}\bra{k}\to\rho_k=\sum_{l=0}^k{k \choose l}\eta^l (1-\eta)^{k-l}\ket{l}\bra{l},
\end{equation}
which introduces an additional contribution in \eqnref{eq:class_capacity}, reducing the information rate and changing the optimal prior distribution in \eqnref{eq:Fock_id}. Despite of the above issues the PNR measurement remains optimal for the Fock states ensemble, since they carry information encoded only in the number of photons. On the other hand a practical POVM saturating the classical capacity bound in \eqnref{eq:ideal} for the coherent states ensemble still remains unknown.

A number of early results in quantum communication theory considered communication through a lossy channel with the Fock states ensemble \cite{Gordon1962, Bowen1968}. In particular, Bowen \cite{Bowen1968} identified approximate expressions for the capacity in the large output number of photons regime $\eta\bar{n}\gg 1$ and the photon starved regime $\eta\bar{n}\ll 1$. In the latter case it was found that the capacity asymptotically behaves like the ultimate bound \eqnref{eq:ideal} $\sim \eta\bar{n}\log_2 \eta\bar{n}$ whereas in the former scenario one obtains
\begin{equation}\label{eq:limit_Bowen}
C_\textrm{Fock}\overset{\eta\bar{n}\gg 1}{\approx} \frac{1}{2}\left[\log_2\eta\bar{n}+\log_2\left(\frac{e}{\pi}\frac{1}{1-\eta}\right)\right].
\end{equation}
Crucially, by \eqnref{eq:limit_Bowen}, the highest information rate achievable with the help of Fock states ensembles asymptotically attains in the leading order only half of the classical capacity of the lossy channel \eqnref{eq:ideal}, irrespectively of $\eta<1$. Note that for the ideal lossless channel, $\eta=1$, the approximation in \eqnref{eq:limit_Bowen} diverges, which is consistent with the fact that the classical capacity in \eqnref{eq:ideal}, saturable with Fock states in the lossless case, has a different asymptotic behavior for large average number of photons. An analysis of the capacity of a communication protocol employing pulse position and pulse amplitude modulations with Fock states was conducted in \cite{Kanaya1989}.

The capacity attainable with the coherent states ensemble and PNR measurement has been extensively studied \cite{Gordon1962, Shamai1990, Martinez2007, Lapidoth2009, Lapidoth2011, Wang2014, Cao2014, Cao2014a, Cheraghchi2019}. In such a scenario, the conditional probability distribution of the detected number of photons depending on the amplitude is given by a Poissonian distribution
\begin{equation}\label{eq:prob_poiss}
p(l|\alpha)=e^{-\eta|\alpha|^2}\frac{\left(\eta|\alpha|^2\right)^l}{l!},
\end{equation}
and hence the resulting classical communication channel is usually called the discrete-time Poisson channel. It was shown \cite{Gordon1962, Martinez2007, Cheraghchi2019} that for such a scenario a non-Gaussian ensemble of coherent states for a large average output number of photons offers an information rate exhibiting a similar behavior as in \eqnref{eq:limit_Bowen}
\begin{equation}\label{eq:Gordon}
R\approx \frac{1}{2}\log_2\eta\bar{n},
\end{equation}
with the exact equality holding in the asymptotic scenario. Similarly, in the opposite regime of weak signals $\eta\bar{n}\ll 1$ one obtains in the first order rate scaling as $\sim \eta\bar{n}\log_2 \eta\bar{n}$, the same as the ultimate bound \eqnref{eq:ideal} and as with the Fock states ensemble. It seems therefore that when one considers measurements of energy at the output both ensembles offer close performance and it is the ability to encode information in optical phase that makes coherent states more robust if one allows for more general detection methods. Note, however, that capacities under the assumption of PNR measurement for both ensembles are known only approximately and in certain regimes. In general, only various upper and lower bounds are known, most notably those derived for coherent states in \cite{Martinez2007, Cheraghchi2019}.

An important point of reference for communication with coherent states is the capacity attainable with quadrature detection. The paradigmatic examples in this scenario are homodyne and heterodyne measurements in which respectively one or two orthogonal quadratures of light are measured. Both methods are limited by the quantum shot noise equal to $1/2$ per quadrature. Capacities achievable by these methods are given by the celebrated Shannon-Hartley theorem \cite{Shannon1949} as
\begin{IEEEeqnarray}{rCl}
C_{\textrm{hom}} &= &\frac{1}{2}\log_2\left(1+4\eta\bar{n}\right),\\
C_{\textrm{het}} &= &\log_2\left(1+\eta\bar{n}\right),
\end{IEEEeqnarray}
respectively. Importantly in the regime of large average received number of photons $\eta\bar{n}$ heterodyne detection saturates the quantum limit in \eqnref{eq:ideal} up to $1$ nat difference, whereas the homodyne detection, similarly as the PNR measurement, attains just a half of the bound $C_\textrm{hom}\approx \frac{1}{2}\log_2\eta\bar{n}$. On the other hand in the opposite regime of small received average number of photons it is the homodyne detection that outperforms the heterodyne measurement by a factor of $2$, offering capacity $C_\textrm{hom}\approx2\eta\bar{n}$ as opposed to $C_\textrm{het}\approx \eta\bar{n}$. Note that in this regime both methods do not attain the classical capacity bound \eqnref{eq:ideal}.

\section{Blahut-Arimoto Algorithm}

Our main goal is to calculate the capacities attainable with the Fock and coherent states ensembles over the lossy channel with PNR detection performed at the output. In order to find the capacity offered by the former ensemble at the input one needs to solve \eqnref{eq:capacity} for a classical information channel given by a conditional probability distribution evaluated through the Born rule on a state in \eqnref{eq:fock_lossy}
\begin{equation}\label{eq:probability}
p(l|k)=\bra{l}\rho_k\ket{l}={k \choose l}\eta^l(1-\eta)^{k-l}.
\end{equation}
Therefore one needs to optimize the mutual information for this channel over a discrete prior probability distribution $p(k)$ with a constraint on the average input signal number of photons $\sum_k p(k) k=\bar{n}$. Note that the input and output of the classical channel defined in \eqnref{eq:probability} are discrete. Despite this fact, unfortunately, optimization cannot be done analytically and one has to rely on numerical means.

We solve the problem of maximizing the rate by employing the Blahut-Arimoto algorithm \cite{Blahut1972, Arimoto1972}, which is designed to compute the capacity of memoryless, discrete channels. The algorithm in each iteration provides a subsequent approximation of the actual value of the capacity and it is guaranteed to converge. Below we present a modified version of the algorithm which allows to apply arbitrary constraints on the prior distribution \cite{Blahut1972}.

First of all, let us write the expression for the mutual information \eqnref{eq:mutual} in an alternative form
\begin{equation}
I(X,Y)=H(X)-H(X|Y).
\end{equation}
The conditional entropy $H(X|Y)$ is equal to
\begin{equation}
H(X|Y) = -{\sum_{x, y} p(x) p(y|x) \log_2 p(x|y)},
\label{eq:cond_entropy}
\end{equation}
where we have used the Bayes' formula $p(x|y)=p(x) p(y|x)/p(y)$. The conditional probability $p(x|y)$ under the logarithm can be interpreted as an example of a stochastic matrix $\Phi_{x|y}=p(x|y)$. A stochastic matrix is defined to be any matrix composed of nonnegative entries $\Phi_{x|y}\geq 0$ with columns that sum up to $1$, i.e., $\sum_x \Phi_{x|y}=1$ for any $y$. The expression in \eqnref{eq:cond_entropy} can be generalized for an arbitrary stochastic matrix $\Phi$ as
\begin{equation}
    J(X|Y; \Phi) = -{\sum_{x, y} p(x) p(y|x) \log_2 \Phi_{x|y}}.
\end{equation}
One can show that the conditional entropy in \eqnref{eq:cond_entropy} is the minimal value of $J(X|Y;\Phi)$
\begin{equation}
    J(X|Y; \Phi) \geq H(X|Y),
\label{eq:J_H_inequality}
\end{equation}
with equality holding when $\Phi$ is of the form reproducing \eqnref{eq:cond_entropy}
\begin{equation}
    \Phi_{x|y} = p(x|y)=\frac{p(y|x) p(x)}{\sum_{x'} p(y|x') p(x')}.
\end{equation}
This observation allows one to write the capacity in \eqnref{eq:capacity} as
\begin{equation}\label{eq:capacity_Blahut}
    C = \max_{\{p(x)\}} \max_\Phi \;\left[ H(X) - J(X|Y; \Phi)\right].
\end{equation}
In order to solve \eqnref{eq:capacity_Blahut} one can perform alternating optimization with respect to $\Phi$ and $p(x)$. Constraints on the input signal may be included in the above optimization using the Lagrange multipliers method by adding a term $\sum_{i=1}^n\lambda_i \left(\sum_x p(x) f_i(x)-K_i\right)$, where $n$ is the number of constraints, $K_i$ denote the constrained values and $f_i(x)$ are functions defining the constraints. Out of these one may single out the constraint $\mu\left(\sum_x p(x) - 1\right)$ which ensures that the probabilities sum up to 1 and always has to be satisfied. The exact algorithm is shown in Algorithm~\ref{con-BA}.

\begin{figure}[t!]
\begin{algorithm}[H]
\floatname{algorithm}{Algorithm}
\caption{Blahut-Arimoto Algorithm With Constraints $\sum_x p(x) f_i(x)=K_i$ on the Input Distribution}
\label{con-BA}
\begin{algorithmic}[1]
\State Choose an initial prior distribution $\{p^{(0)}\}$.
\State In step $t$, maximize $H(\{p^{(t)}\}) - J(X|Y; \Phi^{(t)}) + \sum_{i=1}^n\lambda_i \left[\sum_x p^{(t)}(x) f_i(x) - K_i\right] + \mu\left[\sum_x p^{(t)}(x) - 1\right]$ with respect to $\Phi^{(t)}$, keeping $\{p^{(t)}\}$ constant. The result is $\Phi^{(t+1)}_{x|y} = p(y|x) p^{(t)}(x) / \left[\sum_{x'} p(y|x') p^{(t)}(x')\right]$ according to (\ref{eq:J_H_inequality}).
\State Maximize $H(\{p^{(t)}\}) - J(X|Y; \Phi^{(t+1)}) + \sum_{i=1}^n\lambda_i \left[\sum_x p^{(t)}(x) f_i(x) - K_i\right] + \mu\left[\sum_x p^{(t)}(x) - 1\right]$ with respect to $\{p^{(t)}\}$, keeping $\Phi^{(t+1)}$ constant. The result is $p^{(t+1)}(x) = r^{(t)}(x) / \sum_{x'} r^{(t)}(x')$, where $r^{(t)}(x) = \exp \left[ \sum_i \lambda_i f_i(x) + \sum_y p(y|x) \log \Phi^{(t+1)}_{x|y} \right]$.
\State Find $\lambda_i$ from the constraints $\sum_x p(x) f_i(x) = K_i$ and use them in the formula from step 3 in order to obtain $p^{(t+1)}(x)$ for all $x$.
\State Repeat 2--4 until convergence.
\end{algorithmic}
\end{algorithm} 
\end{figure}
Optimizing the information rate for the discrete-time Poisson channel, i.e., the coherent states ensemble measured with PNR detector is a qualitatively different problem. This is because the input random variable, which in this case is the amplitude of the state, is a continuous random variable and one should in principle optimize over a continuous distribution $\{p(x)\}$ in \eqnref{eq:capacity_Blahut}. This is a considerably more involved task as, in order to use a Blahut-Arimoto-like algorithm, one has to discretize the input alphabet and model the prior probability distribution as a collection of delta functions located at points that have to be optimized. In order to do so we used a continuous version of the Blahut-Arimoto algorithm described in \cite[Sec. 7.3]{Dauwels2006} modified such that it includes the constraint on the average number of photons. The outline of the algorithm is presented in Algorithm~\ref{cont-con-BA} while the details are described in Appendix~\ref{appBA}.

\begin{figure}[t!]
\begin{algorithm}[H]
\floatname{algorithm}{Algorithm}
\caption{Blahut-Arimoto algorithm for continuous input alphabets}
\label{cont-con-BA}
\begin{algorithmic}[1]
\State Choose an initial $n$-point sample $\boldsymbol{x}^{(0)}=(x_1^{(0)},\dots,x_n^{(0)})$ of input symbols which discretizes the continuous input alphabet. The size $n$ of the sample should be chosen large to well approximate the continuous distribution.
\State In step $t$ perform a discrete version of the Blahut-Arimoto algorithm described in Algorithm~\ref{con-BA}. The result is the optimal prior probability distribution $\boldsymbol{p}^{(t)}=(p_1^{(t)},\dots,p_n^{(t)})$ for the sample $\boldsymbol{x}^{(t)}$.
\State Update the sample $\boldsymbol{x}^{(t)} \to \boldsymbol{x}^{(t+1)}$ keeping $\boldsymbol{p}^{(t)}$ constant in a way which increases the mutual information but also maintains the constraint placed on the input, e.g., by a projected gradient descent method:
	\begin{algsubstates}
        \State Calculate $\boldsymbol{g}^{(t)}=(g_1^{(t)},\dots g_n^{(t)})$ where $g_i^{(t)}=\frac{\partial}{\partial x_i^{(t)}}\KLdiv{p(y|x_i^{(t)})}{p^{(t)}(y)}$ and $\KLdiv{p(y|x)}{p(y)}$ is the Kullback-Leibler divergence between distributions $p(y|x)$ and $p(y)$. $p^{(t)}_i$ are treated as constant values independent of $x_i^{(t)}$ and $p^{(t)}(y)=\sum_{i=1}^n p(y|x_i^{(t)})p^{(t)}_i$.
        \State Normalize $\tilde{x}_i^{(t)}=\sqrt{p^{(t)}_i}x_i^{(t)}$ and $\tilde{g}_i^{(t)}=\sqrt{p^{(t)}_i}g_i^{(t)}$ for $i=1,\ldots,n$.
        \State Calculate $\boldsymbol{\tilde{n}}^{(t)}=\bar{n}\tilde{\boldsymbol{g}}_\perp^{(t)}/|\tilde{\boldsymbol{g}}_\perp^{(t)}|$, where $\boldsymbol{\tilde{g}}_\perp^{(t)} = \boldsymbol{\tilde{g}}^{(t)} - \left( \boldsymbol{\tilde{g}}^{(t)} \cdot \boldsymbol{\hat{\tilde{x}}}^{(t)} \right ) \boldsymbol{\hat{\tilde{x}}}^{(t)}$ and $\boldsymbol{\hat{\tilde{x}}}^{(t)}=\boldsymbol{\tilde{x}}^{(t)}/|\boldsymbol{\tilde{x}}^{(t)}|$.
        \State Set $\boldsymbol{x}^{(t+1)}(\phi) = (\boldsymbol{\tilde{x}}^{(t)} \cos \phi + \boldsymbol{\tilde{n}}^{(t)} \sin \phi)/\sqrt{p^{(t)}_i}$.
        \State Optimize mutual information $I(\boldsymbol{x}^{(t+1)}(\phi),\boldsymbol{p}^{(t)})$ over $\phi$, the result is the optimal value $\phi^{*}$.
        \State Update $\boldsymbol{x}^{(t+1)}=\boldsymbol{x}^{(t+1)}(\phi^{*})$.
	\end{algsubstates}
\State Repeat 2--3 until convergence.
\end{algorithmic}
\end{algorithm} 
\end{figure}

\section{Capacity}\label{sec:capacity}

In this section we present capacities attainable by both Fock and coherent states ensembles in the lossy channel detected by the PNR measurement. We performed optimization of the information rate by an appropriate version of the Blahut-Arimoto algorithm in each case, that is the discrete version for the former and continuous for the latter.

\begin{figure}[t!]
\centering
    \includegraphics[width=\linewidth]{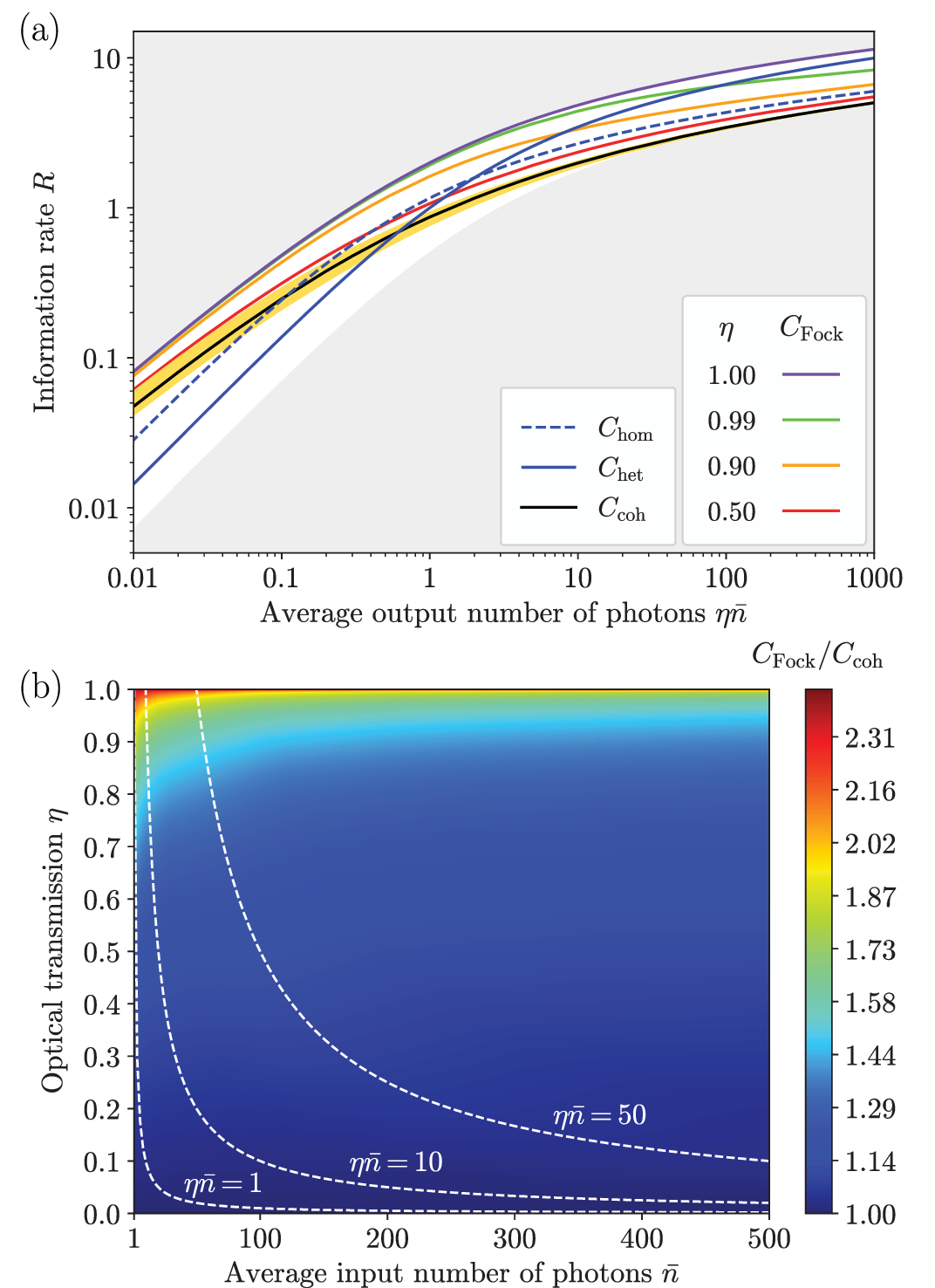}
    \caption{(a) Information rates as a function of the average number of photons at the receiver side. The upper and lower gray regions denote values forbidden by quantum physics and lower than the analytical bound found by Martinez \cite{Martinez2007} respectively. The yellow shaded area indicates a region bounded from above by an upper bound on the coherent states ensemble capacity found in \cite{Cheraghchi2019} and  from below by the result for a Gamma distribution \cite{Martinez2007}. The black solid curve represents capacities attainable with the coherent states ensemble. Capacities for homodyne and heterodyne receivers are shown for comparison. Capacities attainable with the Fock states ensemble are depicted for $\eta = 1$, $0.99$, $0.9$, and $0.5$. Note that Fock states ensembles saturate the classical capacity bound \eqnref{eq:ideal} for $\eta=1$. (b) Ratio of capacities obtained with Fock and coherent states ensembles. White dashed lines depict curves of constant output number of photons $\eta\bar{n}$. Note that $C_{\textrm{Fock}}/C_{\textrm{coh}}$ goes to $1$ as $\bar{n}$ increases and $\eta$ decreases while keeping $\eta\bar{n}$ constant.}
    \label{fig:capacity_comparison}
\end{figure}

It is seen in \figref{fig:capacity_comparison}(a) that irrespectively of the loss parameter, capacities attainable by both Fock and coherent states ensembles, $C_\textrm{Fock}$ and $C_\textrm{coh}$ respectively, converge to the ultimate capacity bound \eqnref{eq:ideal} in the limit of low output average number of photons. In the opposite regime of large output powers, these capacities all converge to a half of the ultimate bound, as indicated in \eqnref{eq:limit_Bowen} and \eqnref{eq:Gordon}, the same value as the one attained by an optimal ensemble of coherent states measured by single homodyne detection. However, it is seen that for all values of $\eta$ the capacities offered by the Fock states ensemble are higher than those obtained with coherent states. Note that the latter depends only on the actual output power $\eta\bar{n}$ since losses just decrease the amplitudes. The coherent states ensemble capacity slightly outperforms the information rate obtained in \cite{Martinez2007} with a Gamma prior distribution but it does not attain the upper bound derived in \cite{Cheraghchi2019}. The difference between capacities offered by both ensembles is further seen in \figref{fig:capacity_comparison}(b), where their ratio is shown. It is seen that the Fock states ensemble offers an almost two-fold advantage in the information rate for links with small losses and moderate signal strength, and the advantage is larger than $1.4$ for $\eta\geq 0.9$. Importantly, in the regime of moderate output number of photons, PNR communication with Fock states outperforms also the one attained by the heterodyne detection of the Gaussian coherent states ensemble, a protocol saturating the classical capacity bound in the large signal strength regime.

\begin{figure}[!t]
\centering
     \includegraphics[width=\linewidth]{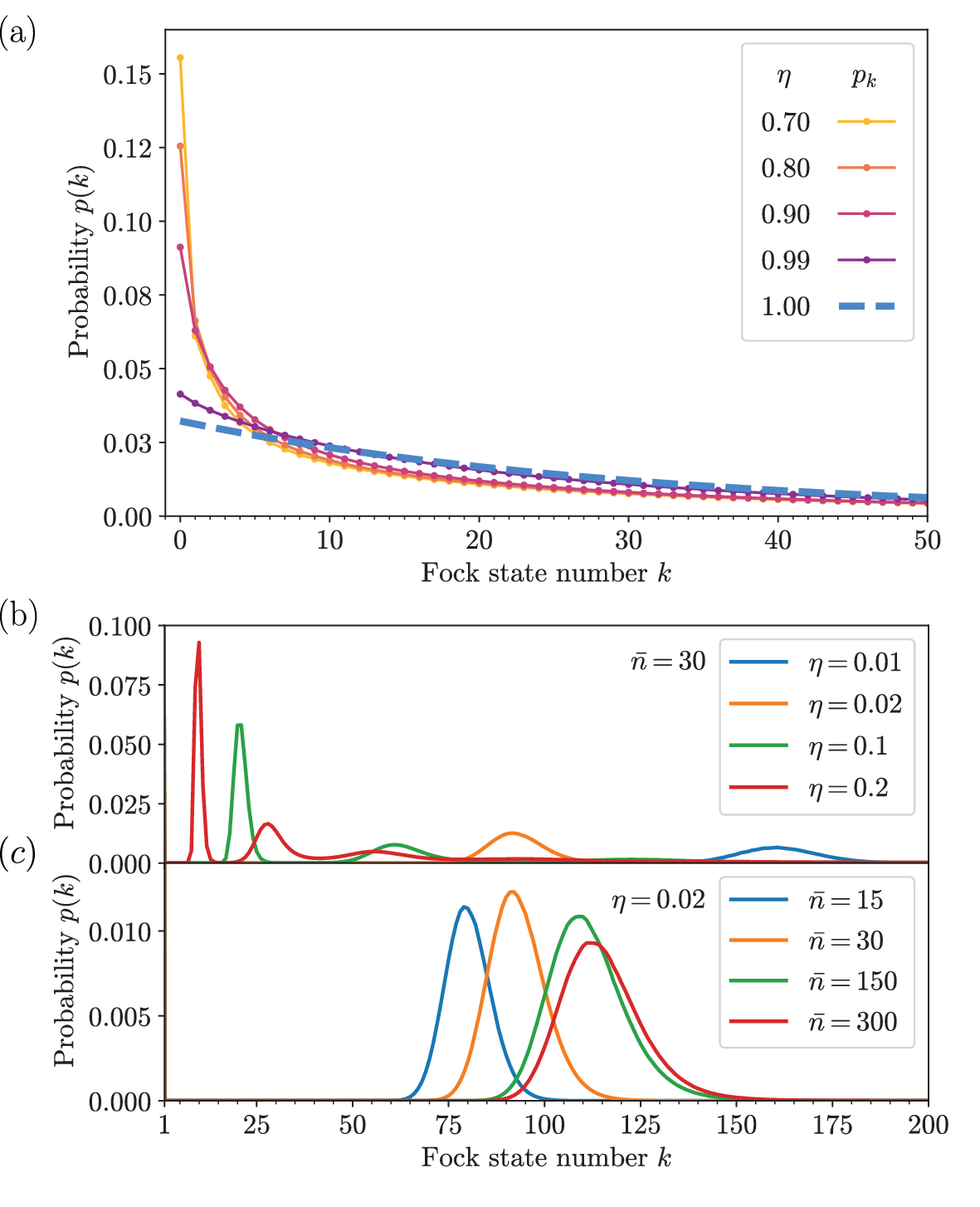}
     \caption{Optimal prior probability distribution $p(k)$ for lossy communication with Fock states ensembles with an average number of photons at the input $\bar{n}=30$ for various levels of losses (a), (b). In (b) the large vacuum state probability $p(0)$ is omitted in order to visualize the behavior for higher Fock state numbers. Prior distribution for a fixed value of losses $\eta=0.02$ and various input average numbers of photons is presented in (c), with omission of large $p(0)$. The curves of respective colors in (b) and (c) correspond to the same value of the output average number of photons $\eta\bar{n}$.}
    \label{fig:probability}
\end{figure}

The optimal prior probability distributions for a particular value of the input average number of photons $\bar{n}=30$ and a few different values of the transmission coefficient $\eta$ are shown in \figref{fig:probability}(a). It is seen that by lowering the transmission, the probability $p(0)$ of sending zero photons raises rapidly and the distribution becomes more peaked around this point. Interestingly, for low transmission the tail of the probability distribution exhibits a few local maxima, as shown in \figref{fig:probability}(b). In the regime of very low transmission coefficients, there is only one such maximum, except of $p(0)$, and its location moves to the higher number of photons the lower $\eta$ one considers. This indicates that in this regime it is optimal to use a version of a generalized on-off keying modulation in which one sends either a vacuum state (empty pulse) or a high-order Fock state \cite{Guha2011, Jarzyna2015, Chung2016}. Importantly, note that the prior distribution is not a function of just the output average number of photons $\eta\bar{n}$. This is seen in \figref{fig:probability}(c) in which the optimal prior distribution is shown for a fixed $\eta$ and a few different average input numbers of photons. It is seen that the resulting optimal prior distributions are different from the ones presented in \figref{fig:probability}(b), even though $\eta\bar{n}$ is the same for respective curves. Despite this difference, however, $p(k)$ still exhibits a single local maximum other than $p(0)$. Note, that by lowering the average number of photons, location of the maximum moves closer to $k=0$.

\section{Poisson channel}

The capacity advantage obtained in Section~\ref{sec:capacity} for Fock states over the coherent ensemble may be easily understood by using the data processing inequality \cite{CoverThomas}. This is because the probability of detecting $l$ photons when the received light was in a state $\ket{\sqrt{\eta}\alpha}$ is equal to
\begin{equation}
p(l|\alpha)=e^{-\eta|\alpha|^2}\frac{(\eta|\alpha|^2)^l}{l!}=\sum_{k=l}^\infty e^{-|\alpha|^2}\frac{|\alpha|^{2k}}{k!}p(l|k),
\end{equation}
where $p(l|k)$ is the probability of detecting $l$ photons when the input was in a Fock state $\ket{k}$, \eqnref{eq:fock_lossy}. Therefore $p(l|\alpha)$ is obtained by mixing respective probabilities for the Fock states ensemble. This means that the mutual information attainable with this ensemble cannot be smaller than what is achievable for coherent states i.e., the capacity of the Poisson channel.

Interestingly, the capacity for the Fock states ensemble in the limit $\eta\to 0$ while keeping $\eta\bar{n}$ constant is equal to the one for the Poisson channel, i.e., attained by coherent states which is seen in \figref{fig:capacity_comparison}(b). This can be rigorously explained with a following argument. For a given prior probability distribution and transmission coefficient $\eta_0$, the average number of photons at the output is equal to $\eta_0\bar{n}_0=\eta_0\sum_{k=0}^\infty p(k) k=\sum_{k=1}^\infty c_k p(k)$, where $c_k=\eta_0 k$. Let us write $c_k=\eta f_k$, where $f_k=k\eta_0/\eta$. If one fixes all $c_k$, then for $\eta\to0$ one has $f_k\to \infty$. Therefore, for very small $\eta$ one also can approximate all $f_k$ by natural numbers. Taking then an ensemble of Fock states with the same probabilities $p(k)$ and $\left|f_k\right\rangle$ instead of $\ket{k}$ one obtains the conditional probability distribution in \eqnref{eq:probability} in the limit $\eta\to 0$ as
\begin{equation}
p(l|f_k)=e^{-c_k}\frac{c_k^l}{l!},
\end{equation}
which is the Poisson distribution one would obtain for a discrete ensemble of coherent states with amplitudes satisfying $\eta_0|\alpha_k|^2=c_k$ in \eqnref{eq:prob_poiss}. Note that the average number of photons at the output for this scenario is still equal to $\eta_0\bar{n}_0$. This means that by optimizing the mutual information for the Fock states ensemble $\{p(k), \left|f_k\right\rangle\}$ over $p(k)$ with a constraint $\eta\sum_k f_k p(k) =\eta_0 \bar{n}_0$ and going with $\eta\to 0$ one obtains the same result as by optimizing the mutual information for the Poisson channel over $p(k)$ and $c_k$ with a constraint $\sum_k p(k) c_k =\eta_0 \bar{n}_0$. Since, as shown above, the Fock states ensemble capacity cannot be lower than what is achievable with coherent states with the same output average number of photons, the capacities attainable by both ensembles are equal to each other in the limit $\eta\to 0$ while keeping the output average number of photons $\eta_0\bar{n}_0$ constant. Note that this means that not only the Poisson channel capacity can be obtained as a limiting case of the result for Fock states but also the optimal prior distributions for both cases are the same and one obtains optimal values of coherent states amplitudes as $c_k\to\eta f_k$ for $\eta\to 0$. This agrees with observations made in \cite{Cao2014, Cao2014a} in that capacity achieving distributions for the Poisson channel exhibit similar behavior as the one described in Section~\ref{sec:capacity} for the Fock state ensemble. Interestingly, this result shows also that, despite the fact that the amplitude $\alpha$ is in principle a continuous parameter, the optimal distributions for the Poisson channel are discrete which was conjectured in \cite{Shamai1990} and then rigorously proved in \cite{Cheraghchi2019}.

\section{Approximation of capacity}

The prior probability distributions optimizing the Fock states ensemble capacity obtained through the Blahut-Arimoto algorithm in Section~\ref{sec:capacity} do not follow any simple closed-form analytical formula. Therefore, in order to approximate the optimal capacity we employ a prior distribution in the form of a negative binomial distribution
\begin{equation}
    p(k) = \frac{\Gamma(k+r)}{k!\Gamma(r)} p^k (1-p)^r,
\label{eq:neg-bin-distr}
\end{equation}
where $p=\bar{n}/(\bar{n}+r)$ is chosen such that the average number of photons is equal to $\langle k \rangle=\bar{n}$ and $r$ is a continuous parameter to be optimized over. Note that the optimal distribution of Fock states in the lossless case \eqnref{eq:Fock_id} is a special case of \eqnref{eq:neg-bin-distr} with $r=1$.

\begin{figure}[!t]
	\centering
    \includegraphics[width=\linewidth]{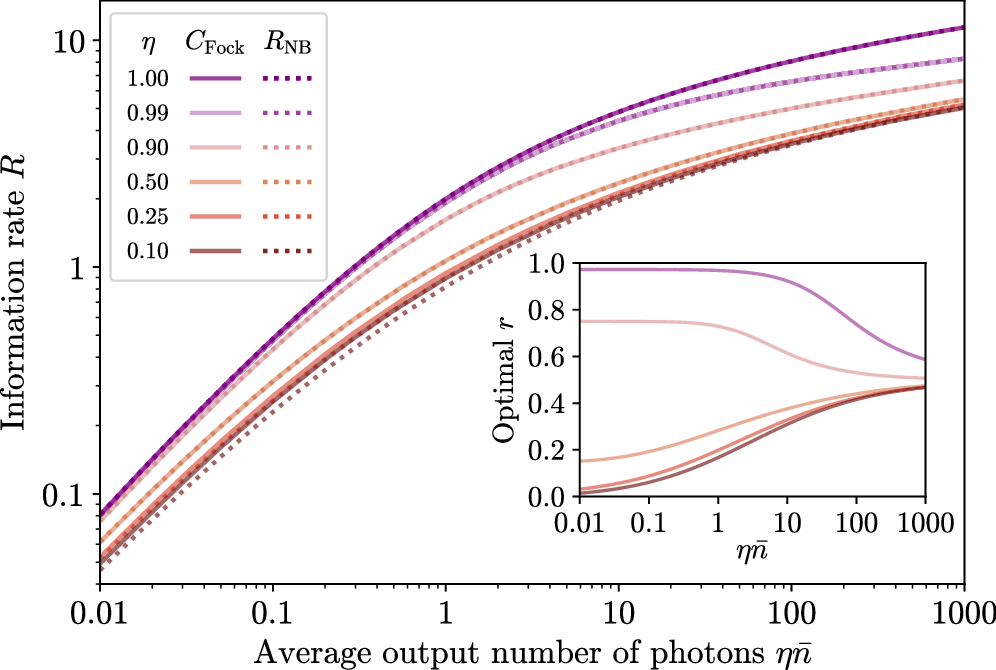}
    \caption{Information rate attainable for a negative binomial prior probability distribution \eqnref{eq:C_negbin} (dotted curves) and capacity $C_\mathrm{Fock}$ attainable in the lossy channel with Fock states ensembles (solid curves). The value of parameter $r$ for negative binomial distribution optimizing the information rate is depicted in the inset as a function of the average number of photons at the output.}
    \label{fig:neg-bin-rates}
\end{figure}

Using integral expressions for the entropies of binomial and negative binomial distributions derived in \cite{Cheraghchi2018} it can be shown (see appendix \ref{app}) that for a given value of $r$ the mutual information for the prior distribution in \eqnref{eq:neg-bin-distr} is equal to 
\begin{IEEEeqnarray}{rCl}
	\IEEEeqnarraymulticol{3}{l}{
        I_\text{NegBin}(X:Y, r) = \frac{1}{\ln 2}\int_{0}^{1}\frac{\text{d}z}{z\ln(1-z) }}\nonumber\\ 
		&&\qquad \times\>\left\{ (1-z)^{r-1} \left[ \left(\frac{r}{r+\eta\bar{n}z} \right)^r + \eta\bar{n}z - 1 \right ]\right. \nonumber\\
		&&\qquad \qquad +\> \left.  \left(\frac{r}{r+(1-\eta)\bar{n}z} \right)^r  - \left(\frac{r}{r+\bar{n}z} \right)^r - \eta\bar{n}z \right\} \label{eq:I_negbin} \nonumber \\
		&&\qquad +\> (\eta\bar{n}+r) h\left(\frac{\eta\bar{n}}{\eta\bar{n}+r}\right) - \bar{n}h(\eta) - \eta\bar{n} \log_2 r, \yesnumber
\end{IEEEeqnarray}
where $h(x) = -x \log_2 x - (1-x)\log_2(1-x)$ is the binary entropy. The information rate attainable with such input distribution reads therefore
\begin{equation}\label{eq:C_negbin}
    R_\text{NB} = \max_r I_\text{NegBin}(X:Y, r)
\end{equation}
and is a lower bound to the true capacity of the Fock states ensemble. It is seen in \figref{fig:neg-bin-rates} that with the help of \eqnref{eq:C_negbin} one is able to achieve information rates very close to the true capacity for $\eta\gtrsim 0.1$. The reason for this is that the negative binomial distribution does not exhibit any sort of multimodal behavior, characteristic for the optimal distribution in the high losses scenario, and shown in \figref{fig:probability}. Nonetheless, for lower transmission the information rate in \eqnref{eq:C_negbin} approximates the lower bound on the capacity for the coherent states ensemble attainable with a Gamma prior distribution found in \cite{Martinez2007}.

\section{Conclusion}

In conclusion we have numerically found optimal prior probability distributions and calculated the capacity for a lossy photon channel with an input Fock states ensemble. We showed that the capacity for such a strategy outperforms the one attainable by an arbitrary coherent states ensemble with PNR detection. The latter scenario is described by a discrete-time Poisson channel, for which the capacity is equal to the one obtained for the Fock states ensemble in the limit of small transmission while keeping the output number of photons constant. Crucially, our result allows not only to evaluate the capacity for the Poisson channel but also identifies the optimal prior probability distribution and amplitudes. 

Our results indicate that in the regime of few photons at the output and low losses the Fock states ensemble offers information rate noticeably higher than the coherent ensemble paired with the PNR detection or quadrature detection. Given the recent progress in both PNR detectors \cite{Fukuda2011, Gerrits2012, Harder2016} and the production of multi photon Fock states \cite{Hofheinz2008, Zhou2012}, they might become relevant in communication scenarios over short distances like, e.g., optical interconnects. A crucial development for this use, however, would be to design PNR detectors characterized by high quantum efficiency which can operate in regular conditions, e.g., without the need of cryogenic cooling. The capacity advantage over the coherent ensemble paired with PNR detection persists, although it is slightly lower, also in the regime of low output number of photons $\eta\bar{n}\leq 1$. In this regime PNR detection can be approximated by a single photon detection which can be realized using existing technology \cite{Hadfield2009}.

 On the other hand, our results show that in the absence of high quantum efficiency photodetectors or in the presence of large losses, the advantage offered by the Fock states ensemble vanishes. This means that in such instances sophisticated quantum states of light are of limited use \cite{Jarzyna2015}. However, because of the connection between the Fock state ensemble and the discrete-time Poisson channel in the limit of low transmission and large average number of photons, our findings may allow to increase the performance of communication in various instances in which one can model transmission by the discrete-time Poisson channel, like for example in long range space optical links \cite{HemmatiBiswasProcIEEE2011, Boroson2018}.

\appendices 

\section{Continuous version of Blahut-Arimoto algorithm}\label{appBA}

To find capacities for continuous alphabets one might implement an extension of the Blahut-Arimoto algorithm described in \cite[Sec. 7.3]{Dauwels2006}. The algorithm aims at approximating the capacity of a continuous channel by discretizing its input alphabet and modeling the prior probability distribution as a collection of delta functions located at points that have to be optimized. Each iteration of the algorithm consists of two parts. First, a discrete version of the Blahut-Arimoto algorithm described in Algorithm~\ref{con-BA} is performed on a discrete sample chosen from the input alphabet in order to find the optimal prior probability distribution of the input symbols in this sample. Next, the symbols of the sample are moved within the input space while their probability distribution is kept fixed, in such a way as to increase the mutual information. Unfortunately, since the optimization over the input sample in the second step is non-convex and may lead to a local maximum of the mutual information, there is no proof of the convergence of this method to the true capacity. As such, the values obtained by this procedure should be interpreted only as lower bounds of the true capacity. Moving the sample symbols can be performed in various ways, like for example by a steepest ascent method applied in the original version of the algorithm in \cite{Dauwels2006}. Note that a special care needs to be taken so that the potential constraints on the input alphabet are satisfied in each step of the procedure. We present below a modified version of the algorithm which accounts for the latter problem for the discrete-time Poisson channel.

Consider a finite discrete $n$-point sample chosen from the input alphabet $\boldsymbol{x} = \{x_1, x_2, \ldots, x_n\}$. In the case of the Poisson channel $x_i$ denotes the amplitude of a coherent state $\ket{x_i}$ and we assume $x_i\in\mathbb{R}$ since PNR detection is insensitive to optical phase. The conditional probability $p(k|x_i)$ of detecting $k$ photons if the state $\ket{x_i}$ was sent is given by the Poisson distribution in \eqnref{eq:prob_poiss}. The first step in the algorithm is to perform a Blahut-Arimoto procedure for the sample $\boldsymbol{x}$. One obtains this way an optimal prior probability distribution $\boldsymbol{p} = \{p_1, p_2, \ldots, p_n\}$ for $\boldsymbol{x}$. This distribution satisfies the constraint on the average input energy
\begin{equation}
\label{eq:coh_en_constraint}
\sum_{i=1}^{n} p_i x_i^2 = \bar{n},
\end{equation}
as the average number of photons in a state $\ket{x_i}$ is equal to $x_i^2$.

The next step is to update the sample $\boldsymbol{x} \to \boldsymbol{x'}$ by moving the elements of $\boldsymbol{x}$ without changing $\boldsymbol{p}$ in a way which increases the mutual information $I(X,Y)$. The latter can be expressed as a function of $\boldsymbol{x}$ and $\boldsymbol{p}$ 
\begin{equation}
\label{eq:mutual_div}
I(X,Y)=I(\boldsymbol{x},\boldsymbol{p}) = \sum_{i=1}^{n} p_i \KLdiv{p(k|x_i)}{p(k)},
\end{equation}
where
\begin{equation}
\KLdiv{p(k|x_i)}{p(k)} =  \sum_k p(k|x_i) \log_2 \left(\frac{p(k|x_i)}{p(k)}\right).
\end{equation}
is the Kullback-Leibler divergence \cite{KullbackLeibler} between probability distributions $p(k|x_i)$ and $p(k)$ and the latter is the distribution of photocounts at the output $p(k)=\sum_{i=1}^np(k|x_i)p_i$. For a fixed probability distribution $\boldsymbol{p}$ one may obtain
\begin{equation}
I(\boldsymbol{x'},\boldsymbol{p}\,) \geq I(\boldsymbol{x},\boldsymbol{p}),
\end{equation}
by moving $\boldsymbol{x} \to \boldsymbol{x'}$ in such a way that for each $i$ the Kullback-Leibler divergence is either increased or stays the same, i.e.,
\begin{equation}
\label{eq:KLdiv_geq}
\KLdiv{p(k|x'_i)}{p(k)} \geq \KLdiv{p(k|x_i)}{p(k)}.
\end{equation}
This task can be performed by means of a steepest ascent method \cite[Sec. 1.2]{Bertsekas1999}, in which one moves $\boldsymbol{x}$ in the direction of the gradient of $\KLdiv{p(k|x_i)}{p(k)}$:
\begin{equation}
\label{eq:x_i_prime}
x_i' = x_i + \lambda \frac{\partial}{\partial x_i} \KLdiv{p(k|x_i)}{p(k)} = x_i + \lambda g_i,
\end{equation}
where $\lambda$ is a (typically small) step size, selected such that \eqref{eq:KLdiv_geq} holds, and we define the gradient $\boldsymbol{g} = \{g_1, g_2, \ldots, g_N \}$ where $g_i= \frac{\partial}{\partial x_i}\KLdiv{p(k|x_i)}{p(k)}$. One has
\begin{IEEEeqnarray}{rCl}
g_i & = & \frac{\partial}{\partial x_i} \KLdiv{p(k|x_i)}{p(k)} \nonumber\\ & = & \sum_{k} \frac{1}{\ln 2} \frac{\partial p(k|x_i)}{\partial x_i} \left ( 1 + \ln \frac{p(k|x_i)}{p(k)} - p_i \frac{p(k|x_i)}{p(k)}\right ), \IEEEeqnarraynumspace
\end{IEEEeqnarray}
where $p_i$ are treated as constants since one wants the probabilities of the sample to remain unchanged in this step of the procedure.

However, note that the update rule defined in \eqnref{eq:x_i_prime} may lead to a violation of the energy constraint~\eqref{eq:coh_en_constraint}, i.e., in general one has
\begin{equation}
\sum_{i=1}^{n} p_i x_i'^2 \neq \sum_{i=1}^{n} p_i x_i^2 = \bar{n}.
\end{equation}
Therefore, if one would naively treat $\boldsymbol{x'}$ obtained from \eqnref{eq:x_i_prime} as a new sample of input symbols, it could result in an unstable and nonconvergent procedure. This is because the Blahut-Arimoto algorithm guarantees an increase of the mutual information in each iteration only if the ensemble from the previous iteration is proper and satisfies the energy constraint~\eqref{eq:coh_en_constraint}. On the other hand, if the Blahut-Arimoto algorithm is run on an improper ensemble violating~\eqref{eq:coh_en_constraint}, such as the one resulting from \eqnref{eq:x_i_prime}, the obtained value of the capacity may be in general lower than the one attained by the ensemble found in the previous step of the full procedure. Consequently, one may end up moving away from the capacity-achieving distribution, as after each application of the symbol update rule and the Blahut-Arimoto algorithm the mutual information may decrease.

In order to remedy the above issue one has to modify the update rule defined in~\eqref{eq:x_i_prime} so that the new sample $\boldsymbol{x'}$ satisfies the energy constraint. This can be accomplished for example by a projected gradient descent method \cite[Sec. 2.3]{Bertsekas1999}, in which one projects the gradient $\boldsymbol{g}$ onto a hypersurface defined by the constraint \eqref{eq:coh_en_constraint} in the space of input symbols. For the discrete-time Poisson channel one can simplify the resulting expressions by rescaling the respective variables by $\sqrt{p_i}$ which is kept constant in this step of the procedure
\begin{equation}
\tilde{x}_i = \sqrt{p_i}\; x_i,\quad \tilde{g}_i = \sqrt{p_i} \; g_i,
\end{equation}
so that the energy constraint reads
\begin{equation}
\sum_{i=1}^{n} \tilde{x}_i^2 = \bar{n},
\end{equation}
and defines an $n$-dimensional constant energy sphere of radius $\sqrt{\bar{n}}$. The rescaled vectors $\boldsymbol{\tilde{x}}$ and $\boldsymbol{\tilde{g}}$ define respectively a point on the constant energy sphere and the direction of the fastest growth of the Kullback-Leibler divergence at this point. To update the sample without leaving the constant energy sphere one may first project the gradient $\boldsymbol{\tilde{g}}$ in the direction perpendicular to $\boldsymbol{\tilde{x}}$ resulting in
\begin{equation}
\boldsymbol{\tilde{g}}_\perp = \boldsymbol{\tilde{g}} - \left( \boldsymbol{\tilde{g}} \cdot \boldsymbol{\hat{\tilde{x}}} \right ) \boldsymbol{\hat{\tilde{x}}}
\end{equation}
where $\boldsymbol{\hat{\tilde{x}}} = \boldsymbol{\tilde{x}} / |\boldsymbol{\tilde{x}}|$ is the normalized rescaled sample vector, and then normalize it such that is has length $\bar{n}$
\begin{equation}
\boldsymbol{\tilde{n}} = \frac{\boldsymbol{\tilde{g}}_\perp}{|\boldsymbol{\tilde{g}}_\perp|} \, \bar{n}.
\end{equation}
One can now define the new update rule as
\begin{equation}
\label{eq:update_projected}
\boldsymbol{\tilde{x}'} = \boldsymbol{\tilde{x}} \cos \phi + \boldsymbol{\tilde{n}} \sin \phi,
\end{equation}
where the parameter $\phi$ can be additionally optimized. It can be seen that $\boldsymbol{\tilde{x}'}$ obtained that way stays on the constant energy sphere, i.e. $|\boldsymbol{\tilde{x}'}| = \bar{n}$ and for small $\phi$ it follows locally the direction of the gradient $\boldsymbol{\tilde{g}}$. Finally, in order to determine the new sample vector $\boldsymbol{x'}$ one needs to rescale back
\begin{equation}
x_i' = \frac{1}{\sqrt{p_i}}\, \tilde{x}_i'
\end{equation}
and optimize the resulting $I(\boldsymbol{x'}, \boldsymbol{p})$ over parameter $\phi$. The latter optimization can be easily performed numerically and is akin to the optimization over parameter $\lambda$ from \eqref{eq:x_i_prime} in a typical steepest ascent algorithm. Note, however, that \eqref{eq:update_projected} actually defines a great circle of the constant energy sphere, being an intersection of the sphere with the plane spanned by vectors $\boldsymbol{\tilde{x}}$ and $\boldsymbol{\tilde{g}}$. Thus, even though the optimal parameter $\phi$ is anticipated to be small as $\boldsymbol{\tilde{g}}$ points in the direction of increasing divergence only in the vicinity of $\boldsymbol{\tilde{x}}$, the update rule \eqnref{eq:update_projected} allows to optimize over the whole range of $\phi \in [0, 2 \pi [$. 

\section{Derivation of \eqnref{eq:I_negbin}}\label{app}

For a negative binomial prior probability distribution of Fock states in \eqnref{eq:neg-bin-distr} the photon number statistics at the output is given by
\begin{equation}
P(l) = \sum_{k=l}^\infty p(l|k) p(k)
\end{equation}
where $p(l|k)$ is the conditional probability in \eqnref{eq:probability}. One obtains
\begin{IEEEeqnarray}{rCl}
	P(l) & = & \sum_{k=l}^\infty {k \choose l}\eta^l(1-\eta)^{k-l} \frac{\Gamma(k+r)}{k!\Gamma(r)} p^k (1-p)^r \nonumber \\ 
	& = & \frac{p^l \eta^l (1-p)^r}{l! \Gamma(r)} \sum_{k=l}^\infty \frac{\Gamma[(k-l)+l+r]}{(k-l)!} [p(1-\eta)]^{k-l}. \IEEEeqnarraynumspace \label{eq:Prob_out}
\end{IEEEeqnarray}
The sum in \eqnref{eq:Prob_out} may be computed using the fact that $\sum_{n=0}^{\infty}\frac{\Gamma(n+m)}{n!}x^{n}=\frac{\Gamma(m)}{(1-x)^{m}}$ and substituting $n = k-l$, $m=l+r$ and $x=p(1-\eta)$. The result reads
\begin{IEEEeqnarray}{rCl}
	P(l) & = & \frac{p^l \eta^l (1-p)^r}{[1-p(1-\eta)]^{l+r}}\frac{\Gamma(l+r)}{l!\, \Gamma(r)} \nonumber \\
	& = & \frac{\Gamma(l+r)}{l! \,\Gamma(r)} P^l (1-P)^r, \label{eq:Prob_out_fin}
\end{IEEEeqnarray}
where $P=p\eta / (1-p(1-\eta)) = \eta\bar{n}/(\eta\bar{n}+r)$. As seen in \eqnref{eq:Prob_out_fin} the probability distribution of the photocounts is also negative binomial with the same value of $r$ as at the input but with a different probability of success $P$.

The mutual information may be written as
\begin{equation}\label{eq:mut_neg}
    I(X:Y,r)= H[P(l)]-\sum_{k=0}^{\infty}p(k) H[p(l|k)],
\end{equation}
where $H[P(l)]$ denotes the entropy of the output negative binomial distribution and $H[p(l|k)]$ is the entropy of the conditional probability distribution in \eqnref{eq:probability}. These entropies can be written using integral formulas derived in \cite{Cheraghchi2018} as
\begin{IEEEeqnarray}{rCl}
	\IEEEeqnarraymulticol{3}{l}{
		H[P(l)] = \frac{r\left(h(P)-P\log_2 r\right)}{1-P} + \frac{1}{\ln 2}\int_{0}^{1} \frac{\text{d}z}{z\ln(1-z)}} \nonumber\\*
		&& \quad \times \> \left[(1-z)^{r-1}-1\right] \left[\left(1+\frac{Pz}{1-P}\right)^{-r}+\frac{Prz}{1-P}-1\right],\yesnumber \IEEEeqnarraynumspace \label{eq:entropy_negbin} \\*\nonumber\\
		&&H[p(l|k)] = kh(\eta) + \frac{1}{\ln 2}\int_{0}^{\infty}\frac{\text{d}t}{t(e^{t}-1)} \nonumber \\*
		&& \quad \times \> \left[\left(1-\eta+\eta e^{-t}\right)^{k} \left(\eta+(1-\eta)e^{-t}\right)^{k}-\left(e^{-t}\right)^{k}-1\right]. \IEEEeqnarraynumspace \label{eq:entropy_conditional}
\end{IEEEeqnarray}
In order to find the conditional entropy in \eqnref{eq:entropy_conditional} let us denote $A_{1}=1-\eta+\eta e^{-t}, A_{2}=\eta+(1-\eta)e^{-t}, A_{3}=e^{-t}$. One has
\begin{IEEEeqnarray}{rCl}
	\IEEEeqnarraymulticol{3}{l}{
		H(Y|X)=\sum_{k=0}^{\infty}p(k) H[p(l|k)] = \sum_{k=0}^{\infty}\frac{\Gamma(k+r)}{k!\Gamma(r)} p^k (1-p)^r}\nonumber \\
		&& \qquad\qquad\;\;\;\, \times \> \left[k h(\eta)+\frac{1}{\ln 2}\int_{0}^{\infty}\frac{A_{1}^{k}+A_{2}^{k}-A_{3}^{k}-1}{t(e^{t}-1)}\text{d}t\right]\nonumber\\
		&&\quad=\frac{(1-p)^{r}}{\Gamma(r)}\left\{ \sum_{k=0}^{\infty}\frac{\Gamma(k+r)}{k!} h(\eta) p^{k}k+\frac{1}{\ln 2}\int_{0}^{\infty}\frac{\text{d}t}{t(e^{t}-1)} \right. \nonumber \\
		&& \qquad \qquad \qquad \;\; \times \left. \left[\sum_{k=0}^{\infty}\frac{\Gamma(k+r)}{n!}p^{k}A_{1}^{k} + \sum_{k=0}^{\infty}\frac{\Gamma(k+r)}{k!}p^{k}A_{2}^{k} \right.\right.\nonumber\\
		&& \qquad \qquad \qquad \qquad - \left. \left. \sum_{k=0}^{\infty}\frac{\Gamma(k+r)}{k!}p^{k}A_{3}^{k}-\sum_{k=0}^{\infty}\frac{\Gamma(k+r)}{k!}p^{k}\right]\right\}\nonumber \\
		&&\quad = \frac{h(\eta)pr}{1-p}+\frac{1}{\ln 2}\int_{0}^{\infty}\frac{\text{d}t}{t(e^{t}-1)} \nonumber\\ 
		&&\qquad \times \> \left\{ \frac{(1-p)^{r}}{(1-pA_{1})^{r}} + \frac{(1-p)^{r}}{(1-pA_{2})^{r}}-\frac{(1-p)^{r}}{(1-pA_{3})^{r}}-1\right\}, 
\end{IEEEeqnarray}
where in the last line we have used the fact that $\sum_{k=0}^{\infty}\frac{\Gamma(k+r)}{k!}p^{k}k=\frac{p}{(1-p)^{r+1}}\Gamma(r+1)$ and $\sum_{k=0}^{\infty}\frac{\Gamma(k+r)!}{k!}p^{k}A^{k}=\frac{\Gamma(r)}{(1-pA)^{r}}$. Changing the integration variable to $z=1-e^{-t}$ one obtains 
\begin{IEEEeqnarray}{rCl}
	\IEEEeqnarraymulticol{3}{l}{
	    H(Y|X) =  \frac{h(\eta)pr}{1-p} - \frac{1}{\ln 2} \int_{0}^{1}\frac{\text{d}z}{z\ln(1-z)}} \nonumber \\ 
	    && \qquad \times \> \left\{ \frac{(1-p)^{r}}{(1-p(1-\eta z))^{r}} + \frac{(1-p)^{r}}{(1-p(1-(1-\eta)z))^{r}} \right. \nonumber \\ 
	    && \qquad\qquad\qquad \qquad\qquad\qquad \;\, \left. - \> \frac{(1-p)^{r}}{(1-p(1-z))^{r}}-1\right\}. \IEEEeqnarraynumspace \label{eq:cond_entropy_final}
\end{IEEEeqnarray}
Plugging the results in \eqnref{eq:entropy_negbin} and \eqnref{eq:cond_entropy_final} to \eqnref{eq:mut_neg} and using the expressions $p=\bar{n}/(\bar{n}+r)$ and $P=\eta\bar{n}/(\eta\bar{n}+r)$ one ends up with the formula in \eqnref{eq:I_negbin}.

\section*{Acknowledgment}

We thank K. Banaszek and R. Demkowicz-Dobrza\'{n}ski for insightful discussions. This work was supported by the Foundation for Polish Science under the "Quantum Optical Technologies" project carried out within the International Research Agendas programme co-financed by the European Union under the European Regional Development Fund.

\ifCLASSOPTIONcaptionsoff
  \newpage
\fi

\bibliographystyle{IEEEtranTCOM}
\bibliography{IEEEabrv,pnr_channel}

\end{document}